# An analytical model to explain under which conditions home batteries create new load peaks in the local grid

Lion Hirth

Hertie School, Berlin, and Neon Neue Energieökonomik, Berlin

hirth@hertie-school.org

**Abstract:** Simulation studies find that price-responsive home batteries and electric vehicles create new load peaks in low-voltage grids; with endogenous wholesale prices most of these peaks disappear. This paper explains both findings with an analytical model of the coincident peak at the transformer. The model is algebraic and solved in closed form, so the parameters that determine whether a new peak emerges and how large it becomes can be identified; they are few. The peak is the background load in the critical hour plus the charging of the local fleet. Under exogenous prices the charging equals the rated power of the fleet. Under endogenous prices the national fleet fills the trough of residual load until the price is flat; the depth of the fill follows from an energy balance in which the price function does not appear. For a V-shaped trough the local charging peak is the share of the grid in national flexibility times the square root of flank gradient times national flexibility. Rated power, price level, network charge and market size do not enter. The model reproduces simulated peaks for a German case within a few percent.

Keywords: analytical model; price feedback; coincident peak; home batteries; electric vehicles; distribution grid

JEL classification: Q41, Q48, L94, D47, C63

# 1. Introduction

Decentralized flexibility is growing fast. Home batteries, electric vehicles and heat pumps are connected to the low-voltage grid. An increasing share of them is dispatched in response to dynamic retail tariffs that pass through the hourly wholesale price. The concern of grid operators is well known: all of these assets respond to the same price signal, so they may act in sync, charging at full rated power in the same cheapest hour of the night. The sum of many small batteries then creates a new load peak on the local transformer at a time that used to be off-peak. Dozens of simulation studies have found such rebound peaks for home batteries and for electric vehicles (Hirth, 2026b, 2026c). If they are right, dynamic pricing trades a benefit for the national power system against a cost in local grids.

Two recent studies have argued that this literature overstates the problem (Hirth, 2026b, 2026c). Almost all simulations treat wholesale prices as exogenous, that is, unaffected by the dispatch they trigger. In reality, charging in the cheapest hours raises the price in those hours, which spreads charging across the night. With prices modeled as equilibrium prices, the simulated rebound peaks shrink from a near doubling of the local import peak to a few percent, or disappear altogether. A third study draws the policy conclusion that the local grid problem of home batteries is small under spot tariffs, and that the usual remedy, time-varying network charges, makes it worse (Hirth, 2026a).

What these studies do not offer is an explanation. They report numbers for one country, one year, one set of household profiles and one battery size, together with many sensitivity runs. They show that the price feedback matters, but not which parameters decide whether a new peak emerges, how large it becomes, and how the answer carries over to other grids, other markets and other assets. A grid planner who wants to know whether her transformer is at risk cannot run a national equilibrium model. A reader who wants to compare the sixty or so published simulations, with their different assumptions on batteries, tariffs, grids and prices, lacks a common frame.

This paper provides that frame. I develop an analytical model of the coincident peak in a low-voltage grid with price-responsive batteries. The model is algebraic and can be solved in closed form. It reduces the peak to the sum of two terms, the background load in the critical hour and the charging of the local battery fleet in that hour; everything else is about determining this second term and the critical hour. The parameters that decide whether a new peak emerges and how large it becomes can be identified, and they are surprisingly few. So can the parameters that do not affect the peak, which is as useful for a grid planner as the first list.

Under exogenous prices, the fleet charges at full rated power in the cheapest hours. The new peak is $C^*_{exo} = B_c + Xp$: the background load $B_c$ in the cheapest hour plus the rated power $Xp$ of the $X$ batteries in the grid. Two parameters decide whether a new peak emerges, the headroom between the old peak and $B_c$, and the rated power. Battery energy, the level and shape of the price curve, the size of the national fleet and the network charge do not enter.

Under endogenous prices, the national fleet of flexible assets fills the trough of national residual load until the price is flat across the charging window. The decisive quantity is the depth of this fill, the distance in gigawatts between the lowest residual load and the level of the price plateau. The local grid charges its share of this depth. For a trough with linear flanks, the local charging peak is $\Lambda_v = s\sqrt{rE}$, where $s$ is the share of the grid in the national flexibility $E$, the energy that all price-responsive assets charge in the trough, and $r$ is the flank gradient of the trough in gigawatts per hour. The new peak is $C^*_{endo} = B_v + s\sqrt{rE}$, with $B_v$ the background load in the hour of the lowest residual load. Four parameters determine the peak, $B_v$, $s$, $E$ and $r$, and the overshoot condition compares $s\sqrt{rE}$ with the headroom $B^{max} - B_v$. Battery energy replaces rated power as the decisive asset parameter; rated power becomes a constraint that does not bind. The peak grows with the square root of local battery energy rather than linearly. The level and the shape of the price function drop out: a flat price means a flat residual load whatever the function, so the water level follows from an energy balance alone.

The model yields a regime boundary. Rated power binds whenever the effective charging duration of the trough is shorter than the full-load duration of the batteries, which happens when national flexibility is small. This explains why exogenous and endogenous simulations coincide at low penetration and diverge above it. It also shows that market size by itself is irrelevant: splitting a market into ten identical zones changes neither the share nor the depth that matter. What matters is the share of a grid in the flexibility of whatever market it belongs to, and the shape of that market's residual load.

I test the model against the simulation results of Hirth (2026b, 2026c). The square-root law reproduces the simulated charging peaks within a few percent across battery penetration, local concentration and battery size. It reproduces the finding that rated power is irrelevant under endogenous prices and that the slope of the price function, the amount of utility-scale flexibility and the level of the network charge hardly matter. Where the model fails, it fails for identifiable reasons: a non-responsive component of the price that synchronizes charging in single hours, and the selection of the critical night by batteries with foresight.

The paper is organized as follows. Section 2 reviews the literature. Section 3 sets up the model. Section 4 derives the general model: the exogenous case, the endogenous case with a trough of arbitrary shape, the question of which flexibility counts, partial price feedback, the regime boundary, the export direction and the relevance for grid planning, and compares the model with the simulations. Section 5 derives a simplified model with a V-shaped trough under conservative assumptions and compares it with the simulations for the German base case. Section 6 concludes.

# 2. Literature review

New load peaks from price-responsive flexibility have been studied almost exclusively by simulation. Gottwalt et al. (2011) find that households with smart appliances may create new demand peaks under day-ahead hourly prices. Flath and Gottwalt (2016) trace such peaks to herding of electric vehicles that respond to the same price. Later studies find the same mechanism in low-voltage grids for electric vehicles, heat pumps and home batteries (Verzijlbergh et al., 2014; Hein et al., 2026; Zadoks et al., 2026), for the plug-in decisions of electric vehicle users (Brockmann et al., 2025) and for thermostatically controlled loads under transactive control (Nazir and Hiskens, 2017). At the system level, the effect is known as an avalanche effect (Kühnbach et al., 2021; Schimeczek et al., 2026). Han et al. (2022) quantify the likelihood of extreme peaks from synchronized household storage, but on simulated time series. Hirth (2026b, 2026c) review these simulations and show that almost all of them take wholesale prices as exogenous. Studies with endogenous prices find smaller peaks (Lilienkamp et al., 2025; Hirth, 2026b, 2026c). Empirical evidence is recent. Samende et al. (2025) document herding in a field demonstration of uncoordinated smart charging. Qiu et al. (2022) find that the load impact of in-home charging depends on the rate and differs from simulated predictions. Ovaere and Vergouwen (2026) find that real-time prices raise household peaks but have limited effects at grid level. Zhou et al. (2022) show from data of 699 residential battery systems that most batteries were full too early to reduce export peaks.

The closest analytical work in energy economics is set at the system level. Peak-load pricing theory already contains a shifting peak. If the peak period bears all capacity costs, demand in the off-peak period can exceed the old peak. Efficient prices then share the capacity costs between both periods (Steiner, 1957). Models of storage with endogenous prices derive the efficient dispatch and capacity of storage from the load or residual load curve (Crampes and Moreaux, 2010; Steffen and Weber, 2013). In a Boiteux-Turvey model with solar generation, competitive equilibria with storage minimize expected system cost (Schmalensee, 2022). Duration-limited storage recovers its capacity costs per peak event rather than per peak hour (Shen et al., 2026). Borenstein and Holland (2005) characterize competitive equilibria in which only some customers face real-time prices. Roozbehani et al. (2012) show that real-time pricing can make prices volatile, increasingly so with the price elasticity of demand relative to that of supply. These models contain the price feedback but no distribution grid.

Control and game theory provide analytical results on decentralized response. Mohsenian-Rad et al. (2010) show that the Nash equilibrium of a game in which households respond to a price that depends on aggregate load minimizes total system cost. Gan et al. (2013) show that decentralized charging of electric vehicles in response to a broadcast price converges to valley filling of the national load curve. Ma et al. (2013) prove that a unique Nash equilibrium exists that comes close to valley filling. De Paola et al. (2016) model arbitrage by many small storage devices as a mean-field game. Liu et al. (2019) derive convergence conditions for decentralized valley filling under distribution network constraints. Phanivong and Callaway (2025) propose a linear price because cost-optimizing loads create congestion for distribution system operators. These models design coordination mechanisms. They do not explain when uncoordinated price response creates a local peak. Game-theoretic models of peak pricing characterize equilibrium peaks under demand charges. Shah and Marden (2025) prove that coincident peak pricing never yields a higher equilibrium peak than anytime peak pricing under perfect information but may do so under uncertain baseline load. Chen et al. (2025) derive closed-form equilibria of coincident peak shaving for two agents and two periods. Zhang et al. (2026) extend this framework but solve it numerically. Anunrojwong et al. (2026) derive battery dispatch with price impact in closed form for linear supply curves. These models concern system peaks or wholesale market power. None of them derives the peak in a local grid.

To my knowledge, there is no algebraic analysis of the question of which parameters decide whether price-responsive flexibility creates a new peak in a distribution grid. The simulations report magnitudes for specific settings without isolating these parameters. The analytical models contain the price feedback or the strategic interaction, but not the local grid in which the peak occurs. The filling of the trough under endogenous prices in this paper is close to the storage models above. What is new is the link from the national trough to the peak of a local grid. This is the gap this paper aims to close.

# 3. Setup

## 3.1. Local grid and batteries

Consider a low-voltage grid behind one transformer. The load of the grid without batteries, net of local PV generation, is the background load $B_t$ in kW, with $t$ the hour. Its annual maximum $B^{max}$ is the old import peak, typically a winter evening. The grid contains $X$ batteries, each with usable energy $e$ in kWh, rated charging and discharging power $p$ in kW and round-trip efficiency $\eta$. The full-load duration $\tau = e/(\eta p)$ is the time a battery needs to charge from empty at rated power; in the base case of Hirth (2026b), 8 kWh, 4 kW and 90 % give 2.2 hours. The local fleet energy is $E_L = Xe$ and the local rated power is $Xp$.

The fleet charges $\Lambda_t$ kW and discharges $\Lambda_t^d$ kW. The coincident load at the transformer is the background load plus charging minus discharging:

$$C_t = B_t + \Lambda_t - \Lambda_t^d \tag{1}$$

The new coincident peak is $C^* = max_t\, C_t$. I call the hour in which it occurs the critical hour. A new peak, or overshoot, means that $C^*$ exceeds $B^{max}$, the peak of the grid without batteries. With this definition the discharge of the fleet at the old peak does not matter: it lowers the old peak, but the question is whether a new one appears above the old level. Section 4.7 discusses the design capacity of the transformer as an alternative reference.

## 3.2. Wholesale market and price feedback

The wholesale price $p_t$ in EUR/MWh is a strictly increasing function $f$ of the national residual load after flexibility $R_t$ in GW, plus a component $\varepsilon_t$ that does not respond to the fleet, following Hirth (2026b):

$$p_t = f(R_t) + \varepsilon_t \tag{2}$$

Residual load after flexibility is national load minus wind and solar generation minus the net discharge of all flexible assets. The residual load before the home battery fleet is $R_t^0$. The price is exogenous if $f$ is flat or if the fleet is too small to move $R_t$; it is endogenous if the fleet moves $R_t$ and $f$ is strictly increasing. The slope $f'$ matters for the wedge-related bounds of Section 4.3 and nowhere else.

The flexible assets in the market are the home batteries of all low-voltage grids plus utility-scale storage, pumped hydro, interconnectors and other flexibility that responds to the same price. Their combined energy charged in the trough of a given night is the national flexibility $E$ in GWh. The share of the local grid in it is $s = E_L/E$. In the base case of Hirth (2026b), 190,000 identical grids with 50 batteries each hold 76 GWh and utility-scale flexibility holds 60 GWh, so $E = 136$ GWh and $s = 2.9 \times 10^{-6}$. Section 4.3 explains when utility-scale flexibility counts and when it does not.

## 3.3. Tariff and arbitrage margin

Households pay the wholesale price plus a network charge $\gamma$ per kWh withdrawn, including taxes and levies. A share $\rho$ of the charge is refunded on energy that is reinjected after storage; self-consumption of stored energy saves the charge in full. A battery that charges at the low price $p_L$ and discharges at the high price $p_H$ earns $\eta(p_H + \rho\gamma - c)$ per kWh charged, where $c$ is the cycling cost per kWh discharged, and pays $p_L + \gamma$. Arbitrage is profitable if

$$p_H - \frac{p_L}{\eta} > \omega, \quad \omega = c + \frac{(1-\eta\rho)\gamma}{\eta} \tag{3}$$

The wedge $\omega$ collects losses, cycling costs and the part of the network charge that is neither refunded nor saved. With $c = 20$ EUR/MWh, $\gamma = 100$ EUR/MWh, full refund and 90 % efficiency, the wedge is 31 EUR/MWh. In 2025 the daily spread of German day-ahead prices exceeded 30 EUR/MWh on 359 of 365 days. I therefore assume throughout that prices fluctuate enough for every battery to charge from the grid at night and to be full at the end of the charging window. This is what distinguishes a spot tariff from a flat tariff, under which no battery charges from the grid. The assumption fails only for tariffs without refund or for high cycling costs; Hirth (2026b) reports smaller peaks in those cases.

Table 1 lists the symbols. The coincident load is denoted $C$ rather than $P$, because $P$ denotes prices in Hirth (2026b, 2026c).

Table 1. Symbols

| Symbol | Meaning | Unit |
|---|---|---|
| $t$ | Hour | h |
| $B_t, B^{max}$ | Background load (local load without batteries, net of PV); its annual maximum, the old import peak | kW |
| $B_c, B_v$ | Background load in the cheapest hour (exogenous) and in the hour of the lowest residual load (endogenous) | kW |
| $\Lambda_t, \Lambda_t^d$ | Charging and discharging of the local fleet | kW |
| $C_t, C^*$ | Coincident load at the transformer; the new coincident peak | kW |
| $H$ | Headroom: old peak minus background load in the critical hour | kW |
| $X, p, e, \eta$ | Number of batteries, rated power, usable energy, round-trip efficiency | –, kW, kWh, – |
| $\tau$ | Full-load duration $e/(\eta p)$ | h |
| $E_L$ | Local fleet energy $Xe$ | kWh |
| $E, s$ | National flexibility charged in the trough; share of the local grid, $E_L/E$ | GWh, – |
| $R_t, R_t^0$ | National residual load after flexibility; before the home battery fleet | GW |
| $R_{min}, D, L$ | Lowest residual load; depth of the fill; plateau level $R_{min} + D$ | GW |
| $V(D), W, W_{eff}$ | Volume of the trough below depth $D$; its width; effective charging duration $V/D$ | GWh, h, h |
| $r$ | Flank gradient of the trough (V-shaped case) | GW/h |
| $\alpha, \kappa$ | Exponent and scale of a power-law trough | –, $GW/h^{\alpha}$ |
| $p_t, f, \varepsilon_t$ | Wholesale price; price function of residual load; non-responsive component | EUR/MWh |
| $p_L, p_H, \omega$ | Low and high price of a cycle; wedge | EUR/MWh |
| $c, \gamma, \rho$ | Cycling cost; network charge including taxes; refund share | EUR/MWh, EUR/MWh, – |
| $\Pi_t, F_t, F^{max}$ | Local PV generation; export flow $\Pi_t - B_t + \Lambda_t^d - \Lambda_t$; old export peak | kW |
| $R_{max}, r_h, D_h, E_h$ | Peak of the residual load hill; its flank gradient; depth of the shave; energy available for discharge | GW, GW/h, GW, GWh |
| $K$ | Design capacity of the transformer, the larger of $B^{max}$ and $F^{max}$ | kW |
| $N$ | Number of identical grids in the simulations of Hirth (2026b) | – |

# 4. The general model

## 4.1. Exogenous prices

If the price path does not respond to the fleet, every battery solves the same problem facing the same prices. Each charges at rated power in the $\tau$ cheapest hours of the night and is full afterwards. The coincidence factor is one; nothing dampens it. The fleet charging in the cheapest hour is the rated power of the grid:

$$\Lambda_t = Xp \quad for\ the\ \tau\ cheapest\ hours \tag{4}$$

The new coincident peak is the background load in the cheapest hour, $B_c$, plus the rated power:

$$C^*_{exo} = B_c + Xp \tag{5}$$

The critical hour is the cheapest hour of the night with the highest background load, over all nights of the year. An overshoot occurs if and only if the rated power exceeds the headroom in that hour:

$$C^*_{exo} > B^{max} \quad \Leftrightarrow \quad Xp > B^{max} - B_c \tag{6}$$

Two factors decide: the headroom $B^{max} - B_c$, which depends on when the cheapest hours fall relative to the local load profile, and the rated power $Xp$. Above the threshold the peak grows by $p$ per additional battery. Battery energy does not enter, except that it sets how long the full-power charging lasts, which matters for thermal limits and for peak measures that average over several hours. The level, spread and functional form of the price curve do not enter either, nor does the size of the national fleet, the amount of utility-scale flexibility or the network charge as long as it is refunded or saved. The same equations describe time-of-use tariffs and time-varying network charges, because a deterministic signal set in advance is a price path that does not respond, and a single grid in a large market, because its fleet is too small to move the price.

## 4.2. Endogenous prices: filling the trough

If the price responds, a battery that charges in the cheapest hour raises the price in that hour and lowers its own margin. In equilibrium, no battery can gain by moving charging to another hour, so the price is flat across the hours in which the fleet charges. Because $f$ is strictly increasing, a flat price means a flat residual load after flexibility: the fleet fills the trough of residual load up to a level $L$ and the residual load after flexibility is $L$ wherever it charges. This is the water level of the trough.

Let $R_{min}$ be the lowest residual load before the fleet and $D = L - R_{min}$ the depth of the fill. The volume of the trough below depth $D$ is

$$V(D) = \int (R_{min} + D - R^0_t)^+ \, dt \tag{7}$$

where the integral runs over the night and the superscript plus denotes the positive part. The fleet charges its entire energy in the trough, so the depth follows from an energy balance that involves no price at all:

$$V(D) = E \quad \Rightarrow \quad D = V^{-1}(E) \tag{8}$$

This is the central result of the model. The price function does not appear. Whether the price rises steeply or gently with residual load changes the price level of the plateau, but not its position in gigawatts, because the plateau is defined by flat residual load and the energy that has to fit below it. The price function returns only in two places: the wedge sets a ceiling on the depth, because the fleet stops charging when the plateau price reaches the high price minus the wedge, and the non-responsive component $\varepsilon_t$ can override the plateau in individual hours. Both are treated in Sections 4.3 and 4.4.

At a flat price every battery is indifferent between the hours of the window. I assume that the aggregate charging is allocated across batteries in proportion to their energy, so that each grid charges in every hour its share $s$ of the national charging $(L - R_t^0)^+$. The local charging peak then occurs in the hour of the lowest residual load and equals the share times the depth:

$$\Lambda_v = s \cdot D = s \cdot V^{-1}(E) \tag{9}$$

An equivalent formulation uses the effective charging duration $W_{eff} = V(D)/D$, the duration a rectangular trough of the same depth and volume would have. Local charging is local energy divided by this duration:

$$\Lambda_v = \frac{E_L}{W_{eff}}, \quad W_{eff} = \frac{V(D)}{D} \tag{10}$$

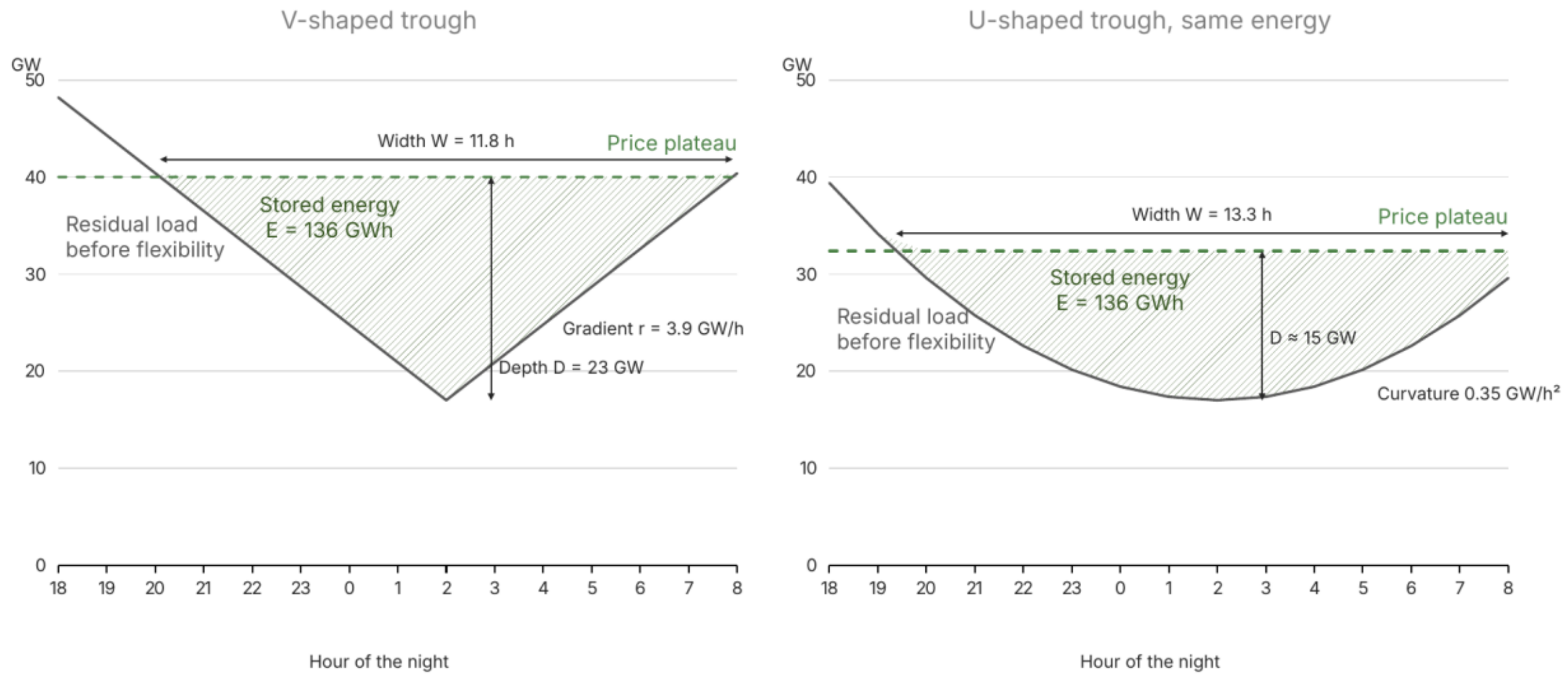

Figure 1. National residual load on the critical night with the stored energy of the national fleet: V-shaped and U-shaped trough with the same energy (stylized, base case parameters of Hirth, 2026b).

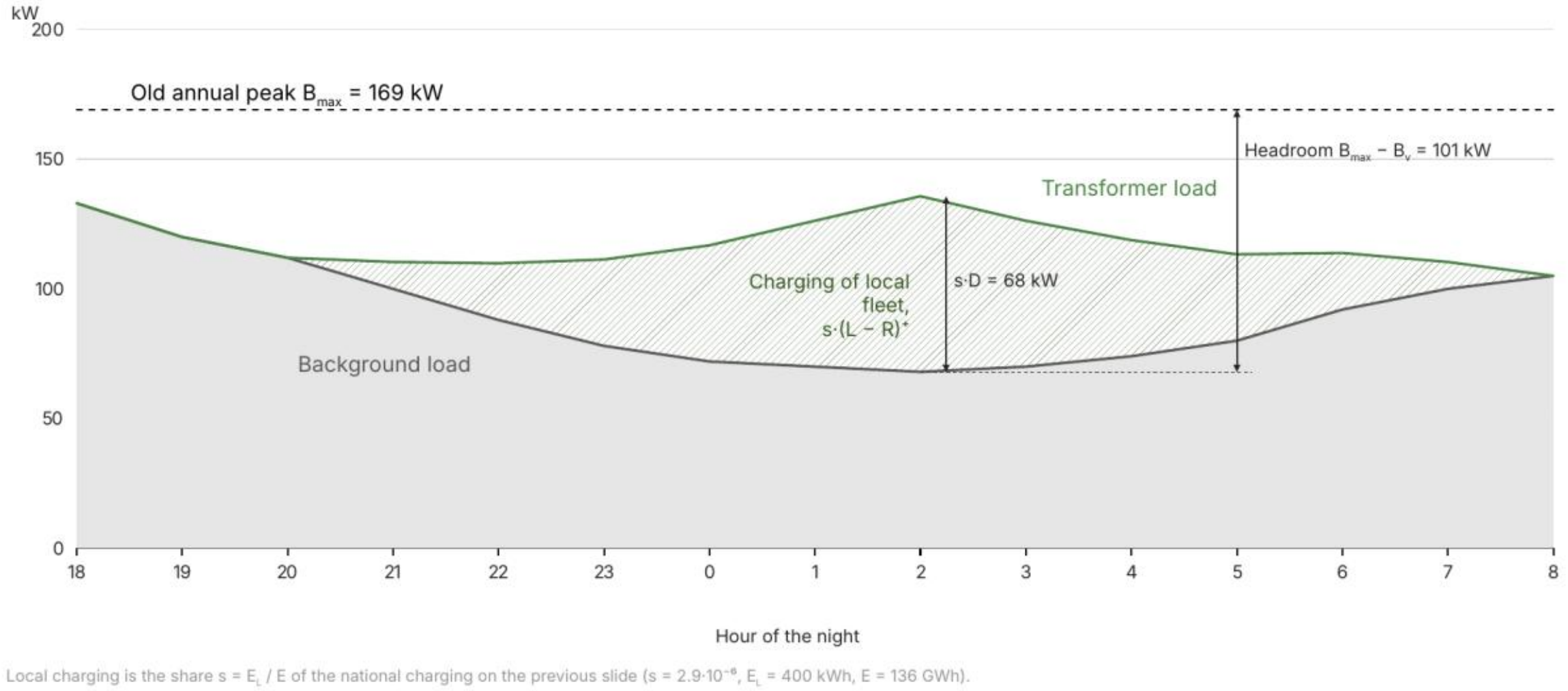

Figure 2. Local import load on the critical night: background load plus the charging of the local fleet, which is the share $s$ of the national charging in Figure 1 (stylized).

The effective charging duration depends on the national flexibility and the shape of the trough alone. The national fleet decides how widely charging is spread; the local grid decides how much energy has to pass through that width. The new coincident peak is the background load in the hour of the lowest residual load, $B_v$, plus the local charging:

$$C^*_{endo} = B_v + s \cdot V^{-1}(E) \quad (11)$$

An overshoot occurs if and only if the share times the depth exceeds the headroom at the lowest residual load:

$$C^*_{endo} > B^{max} \quad \Leftrightarrow \quad s \cdot V^{-1}(E) > B^{max} - B_v \quad (12)$$

Equations (11) and (12) are the result of this paper in its general form. Three quantities determine the new local peak: the background load in the hour of the lowest national residual load, $B_v$; the share of the local fleet in national flexibility, $s$; and the depth of the fill, $D = V^{-1}(E)$, which is set by the national flexibility that has to fit below the plateau and by the volume function of the trough, that is, its shape. The first is a question of timing: a trough at 3 a.m. meets the local load minimum, a trough at 8 p.m. meets the evening peak. The second and the third enter as a product, share times depth, where the depth grows with national flexibility at a rate set by the shape: the more pointed the trough, the less volume fits per gigawatt of depth and the deeper the fill. Nothing else enters. The price function cancels out, as Equation (8) shows. Battery energy enters through $E_L$ and $E$; rated power does not enter as long as the constraint $\Lambda_v \leq Xp$ is slack, which Section 4.5 makes precise. The number of batteries, the efficiency, the network charge and the wedge decide whether this regime applies, not how high the peak is. Where the rest of the national flexibility is located is irrelevant as long as it follows the same arbitrage logic, which Section 4.3 qualifies.

## 4.3. Which flexibility counts

Equation (8) treats all flexible assets alike. They are alike in the trough only if they stop charging at the same plateau price. Each type of storage $i$ charges as long as the plateau price is below its threshold

$$\bar{p}_i = \eta_i\, p_H - \eta_i\, c_i^d - c_i^c - (1 - \eta_i\, \rho_i)\, \gamma_i \tag{13}$$

where $c^d$ and $c^c$ are cycling costs on discharge and charge, and $\gamma_i$ and $\rho_i$ are the network charge and refund share that apply to the asset. In the simulations of Hirth (2026b), home batteries pay cycling costs only on discharge and face the network charge only on their losses, because the refund and the self-consumption saving cancel the rest; utility-scale storage pays cycling costs in both directions and no network charge. With equal efficiency, the threshold of home batteries lies 10 EUR/MWh above that of utility-scale storage: 20 EUR/MWh of charging-side cycling costs in favor of home batteries, minus 10 EUR/MWh of network charge on their losses.

The ordering determines which flexibility belongs in $E$. As the home fleet fills the trough, the plateau price rises. Once it passes the threshold of utility-scale storage, that storage stops charging and the home fleet fills the rest of the trough alone. From then on $E$ contains the home fleet only, and the depth follows from its energy alone. The volume between the two thresholds, the displacement volume, is the price difference divided by the slope of the price function times the width of the window; for the German base case it is about 25 GWh, which the home fleet exceeds above roughly 15 % of households. Below that, both fleets share the trough and the allocation between them is indeterminate, because the price is flat. If the cost ordering were reversed, as it may be for grid-scale batteries with low cycling costs, utility-scale storage would occupy the bottom of the trough first and the home fleet would charge less in the critical hour. Utility-scale storage still affects the home fleet from the other side of the cycle: its discharge lowers the evening price and narrows the spread on which the home fleet bases its charging decision.

The lesson for Equation (8) is simple. The national flexibility that counts is the home fleet plus the part of other flexibility whose threshold lies above the plateau price the home fleet sets. With equal thresholds, everything counts; with a cheaper home fleet of sufficient size, only the home fleet counts.

## 4.4. Partial price feedback

Two mechanisms reintroduce synchronization into the endogenous case. The first is the non-responsive price component $\varepsilon_t$ in Equation (2). If the price in some hour is far below the plateau for reasons unrelated to residual load, every battery charges at rated power in that hour regardless of the feedback. The condition is that the price after full-power charging of the national fleet still lies below the discharge price minus the wedge:

$$f(R_t^0 + NXp) + \varepsilon_t < p_H - \omega \tag{14}$$

with $N$ the number of grids. In the simulations of Hirth (2026b), such hours occur around midday on sunny days with strongly negative $\varepsilon_t$; the fleet then charges at rated power, and the local grid is protected only by its own PV, which masks the withdrawal. The mirror image, a single-hour spike of $\varepsilon_t$ above the plateau, triggers discharge at rated power and matters for the export direction (Section 4.6). The second mechanism is a deterministic component of the retail tariff: time-of-use prices and time-varying network charges are set in advance and do not respond. The retail price is then the sum of a responsive and a non-responsive part, and the fleet behaves as a mixture of Equations (5) and (11). In the simulations of Hirth (2026a), a time-of-use tariff derived from equilibrium prices produces the same peak as exogenous prices, because the deterministic part dominates the small residual variation of the endogenous part.

## 4.5. Regime boundary and market size

Rated power binds when the effective charging duration of the trough is shorter than the full-load duration of the batteries:

$$W_{eff}(E) < \tau = \frac{e}{\eta p} \quad \Rightarrow \quad \Lambda_v = Xp \tag{15}$$

The effective charging duration is set by national flexibility and the shape of the trough. If national flexibility is small, the fleet fills only the bottom of the trough, the effective duration is short, and every battery charges at rated power: the endogenous case reduces to the exogenous one. This is why simulations with exogenous and endogenous prices coincide at low penetration and diverge once the national fleet is large enough to widen the plateau beyond the full-load duration. With 60 GWh of utility-scale flexibility in the trough, the German case is in the energy-constrained regime from the first home battery onward.

Market size by itself does not enter. Suppose a market is split into ten identical zones, each with a tenth of the thermal fleet, a tenth of the residual load and a tenth of the flexibility, and prices clear separately in each zone. The trough in a zone is a tenth of the national trough, so its volume at any depth is a tenth; a tenth of the flexibility fills it to the same depth. The share of a grid in the zone is ten times its share in the nation. Share times depth is unchanged. What changes with zonal pricing is heterogeneity: a grid sees the trough of its region rather than the national average, its share is measured against the flexibility of its region, and its load and PV are more closely correlated with the regional residual load. The extreme case shows that market size does not even fix the sign: a zone that consists of a single distribution grid with its own price fills its own load trough and flattens it, the opposite of a new peak.

## 4.6. Export direction

The export direction mirrors the import direction with two differences: local PV generation $\Pi_t$ drives the flow, and the background load counteracts it. The export flow at the transformer is

$$F_t = \Pi_t - B_t + \Lambda_t^d - \Lambda_t \tag{16}$$

Its maximum without batteries, $F^{max}$, is the old export peak, typically a sunny midday. Charging at midday reduces the export peak; it cannot raise it. A new export peak is a discharge event: the fleet discharges into a price peak in an hour in which the grid already exports or barely imports. The counterpart of the trough is the hill of residual load, the morning or evening peak.

Under exogenous prices the fleet discharges at rated power in the most expensive hours. The new export peak is the background export flow in the price-peak hour plus the rated power, and an overshoot requires that this exceeds the old export peak:

$$F_{exo}^{*} = \Pi_h - B_h + Xp \quad > \quad F^{max} \tag{17}$$

where the subscript $h$ marks the hour of the price peak. The condition has a prerequisite that the import direction lacks: the battery must hold energy at that hour, either from the night before, if the night-to-morning spread exceeds the wedge, or as what remains of the midday charge after the evening discharge.

Under endogenous prices the fleet shaves the hill the way it fills the trough. It discharges until the price is flat across the discharge window, which means flat residual load at a level $L_h$ below the hilltop $R_{max}$. The depth of the shave $D_h = R_{max} - L_h$ follows from the energy available for discharge, $E_h$, which is at most the stored energy times the efficiency:

$$V_h(D_h) = E_h \quad \Rightarrow \quad D_h = V_h^{-1}(E_h), \quad \Lambda_h^d = s \cdot D_h \tag{18}$$

$$F_{endo}^* = \Pi_h - B_h + s \cdot D_h \quad > \quad F^{max} \tag{19}$$

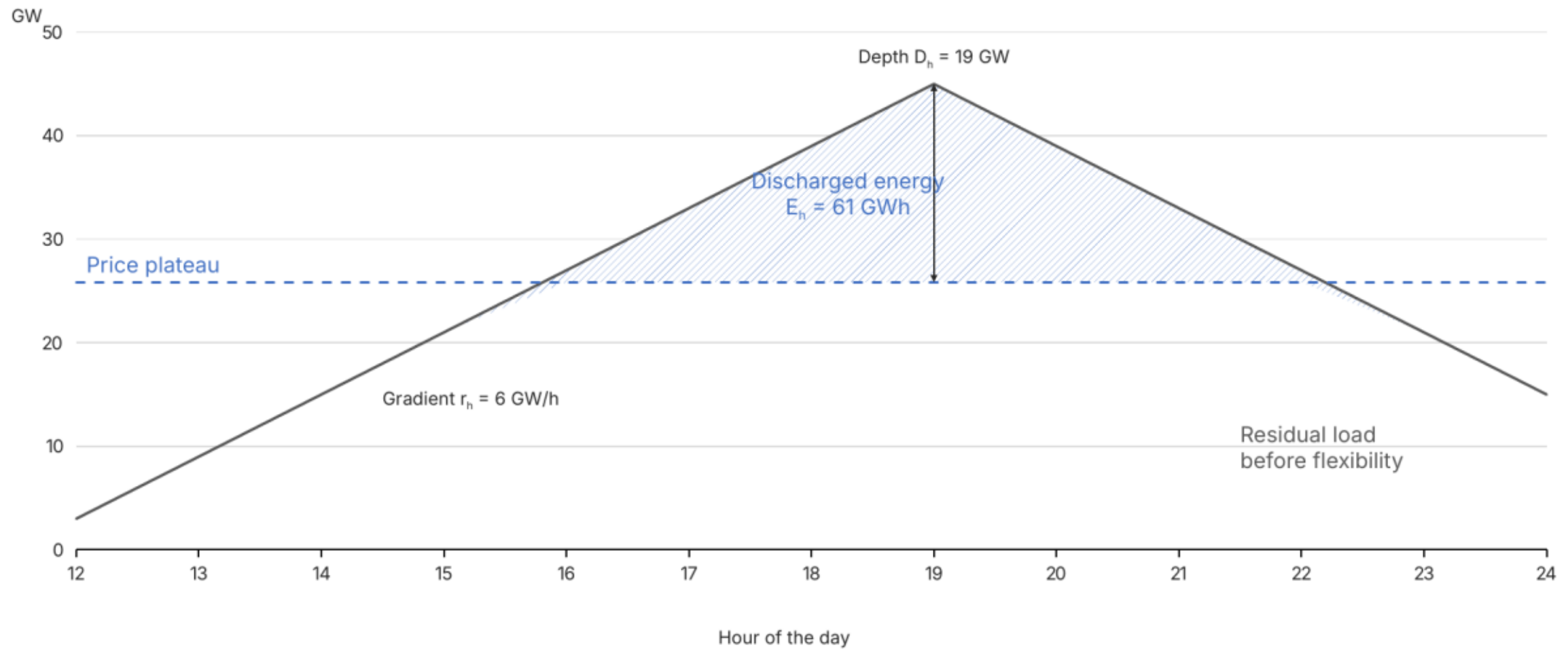

Figure 3. National residual load hill in the evening with the energy discharged by the national fleet (stylized).

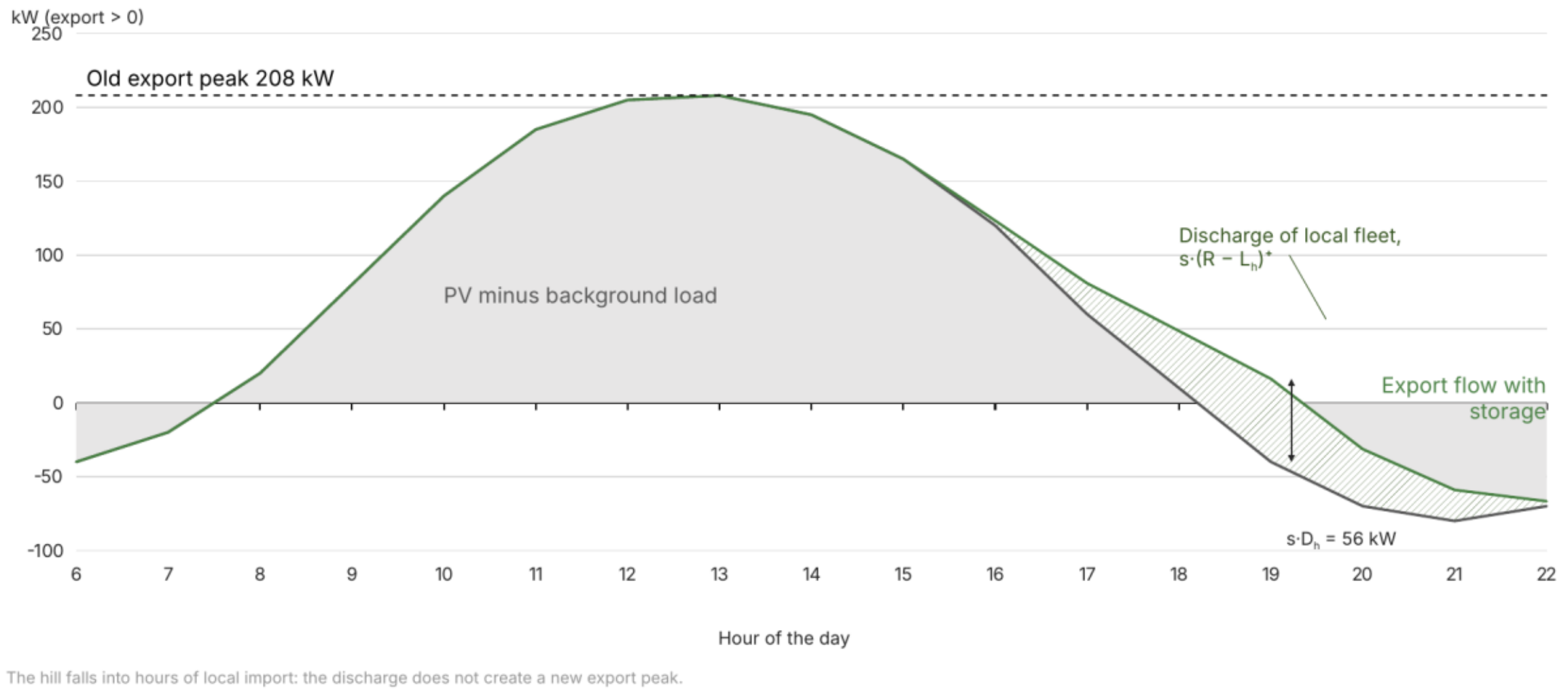

Figure 4. Local export flow: PV minus background load plus the discharge of the local fleet, which is the share $s$ of the national discharge in Figure 3 (stylized).

Two features dampen the export direction relative to the import direction. The hill of residual load lies where national solar generation is low; with correlated weather, local PV is low in the same hours, so the background flow at the hilltop is usually an import and the discharge is absorbed locally. A new export peak requires a hill that reaches into the hours of local surplus, as on summer mornings between 7 and 9 a.m., or weather with low correlation between the grid and the nation. The second feature is the energy available at the hill. In summer the fleet charges at midday and discharges in the evening; at the morning hill a small battery is empty and only a large one still holds energy. The energy available for discharge is therefore a parameter of its own, bounded by battery energy but not equal to it, unlike on a winter night, when the fleet is full at the bottom of the trough.

The charging side at midday deserves a remark. The solar trough is much larger than the fleet: a household with 8 kWp has a daily surplus of 40 to 50 kWh in summer, the battery takes 8. The fleet is full after two hours of an eight-hour surplus, and which two hours it picks is not determined by the trough, because the price is flat across it. The reduction of the export peak by midday charging is therefore small and depends on the tie-breaking rule rather than on the model. The same two hours become a problem in the import direction if local PV is absent while national solar is at its peak. In that case midday is the exogenous regime in the local grid: the background load is the midday load without PV, the headroom to the old import peak is about 110 kW in the base case, and the rated power of 200 kW exceeds it. Even without a non-responsive price component, the solar trough is steeper than the wind trough, and the share times depth at midday is on the order of 145 kW in the base case. Local cloud cover under a national solar peak is thus a case in which the midday charge creates an import peak. The simulations of Hirth (2026b) represent local PV with the reanalysis profile of one grid point near Berlin against the national solar profile, so hours of low correlation between the grid and the nation do occur, but a reanalysis grid cell averages out local cloud fields. How often a single grid sees little PV during a national solar peak is an empirical question that requires metered PV generation data.

## 4.7. Relevance for grid planning

A transformer is rated for the larger of the two directions. In grids with much rooftop PV, the export peak exceeds the import peak, so the design capacity is $K = F^{max}$ and the relevant headroom for a new import peak is $K - B_v$ rather than $B^{max} - B_v$. In the base case of Hirth (2026b), this raises the import headroom from 101 to 140 kW. The rule that follows is simple. Power-constrained events, which are exogenous prices, non-responsive price spikes and weather with low correlation between the grid and the nation, become relevant for the transformer rating as soon as the rated power in the grid exceeds the headroom to the design capacity, in the base case at about 40 batteries. Energy-constrained events, which are endogenous winter nights, become relevant only when share times depth exceeds that headroom, that is, with local concentration or large batteries. Which direction binds first depends on the amount of rooftop PV in the grid: with much PV the design capacity is the export peak, the import headroom is large and the risk comes from discharge into price peaks; with little PV, as in grids serving apartment buildings or grids with electric vehicles but no rooftop solar, the design capacity is the import peak and winter nights and middays matter.

## 4.8. Comparison with simulations

I compare the model with the simulation results of Hirth (2026b) for home batteries, which use 100 households, 50 of them with rooftop PV, in 190,000 identical grids, a cubic price function with a non-responsive component calibrated to German day-ahead prices of 2025, and utility-scale flexibility of 10 GW and 60 GWh. The comparison concerns the peak charging of the local fleet in winter nights, November to February, 6 p.m. to 8 a.m., which is the quantity the model determines. The base case has 50 batteries of 8 kWh and 4 kW.

Under exogenous prices, the cheapest hour of the critical night in the base case is 4 a.m. on 10 January with $B_c = 78$ kW, the old peak is 169 kW and the rated power of 50 batteries is 200 kW. Equation (5) gives 278 kW, the simulation 278 kW. The threshold of Equation (6) is 23 batteries; the simulation finds the import peak above its no-battery level from 15 to 20 % of households, that is 15 to 20 batteries measured on the top 1 % of hours.

Under endogenous prices, the critical night of the simulation is 9 to 10 December. The residual load before utility-scale flexibility falls to 17 GW at 5 a.m. With the home fleet alone in the denominator, as Section 4.3 suggests for the cost ordering of the simulation, the trough of that night holds the 76 GWh of the home fleet at a depth of 17 GW, the effective charging duration is 4.5 hours, and the local charging peak, share times depth, is 91 kW; the simulation shows 93 kW. The background load at the lowest residual load is 68 kW, so Equation (11) predicts a coincident load of 159 kW on that night versus 163 kW in the simulation, both below the old peak of 169 kW. With utility-scale flexibility included in the denominator, 136 GWh fill the trough to a depth of 23 GW, and the model gives 68 kW of charging and a load of 136 kW: the simulated home fleet has displaced the utility-scale storage from the trough, as the cost ordering implies.

The model does not predict which night is critical. Applied night by night with each night's own gradient and a full charge every night, it correlates only weakly with the simulated peaks across nights. Batteries with monthly foresight charge only in the nights with the largest spreads, and the non-responsive component decouples price and residual load in individual hours; on 24 February, for instance, the residual load bottoms out at 8 p.m. at a high local load, which the model flags as dangerous, but the price in that hour is not low and the fleet discharges. The critical night is therefore an input of the model, not an output: the steepest deep trough with a high background load at its bottom.

The simulations also confirm the picture of the export direction. Under exogenous prices, 60 of the 88 hours that form the top 1 % of export hours fall between 5 and 8 a.m., from February to June and in September; the annual export maximum of 234 kW occurs on 17 April at 7 a.m., with a background flow of 34 kW and the fleet discharging its full 200 kW into the morning price peak. Under endogenous prices there are no discharge-driven export peaks in the base case; the top hours shift to 9 and 10 a.m., the PV hours before the fleet starts charging, and lie 15 % below the no-battery level. With large batteries of high rated power, evening export peaks appear even under endogenous prices; they are discharge events in individual hours in which the non-responsive price component dominates, Equation (14) on the discharge side, and they scale with rated power as the exogenous case does.

# 5. A simplified model

The general model needs the volume function of the trough. A trough with linear flanks ("v-shaped") has a closed form and is, for a given volume, a conservative assumption: it is more pointed than a trough with a rounded bottom, so the same energy fills it deeper and the local charging peak is higher. The other assumptions of this section are conservative in the same sense. Every battery is empty at the bottom of the trough and charges its full energy during the night, which overstates the peak in nights with a spread barely above the wedge. The aggregate charging is allocated in proportion to energy, which places the local charging peak at the bottom of the trough rather than spreading it across the plateau. The background load is taken in the critical hour, not on average. In this sense the simplified model is an upper bound, with two exceptions. Counting utility-scale flexibility in the denominator understates the peak whenever the home fleet is cheaper and displaces it, by about a quarter in the base case; the home-fleet-only version is the conservative one and the one the simulation confirms. Neglecting the non-responsive price component understates the peak by about a fifth in the simulation. A user of the simplified model should count the home fleet alone and add a margin for the non-responsive component.

## 5.1. Trough, fill and local peak

To simplify the algebra, I assume the residual load before the fleet to fall linearly to its minimum at $t_0$ and to rise linearly afterwards, with the same gradient $r$ on both flanks; with unequal flanks, $r$ is their harmonic mean:

$$R_t^0 = R_{min} + r \cdot |t - t_0| \tag{20}$$

The volume below depth $D$ is a triangle, and the width of the fill is the base of the triangle:

$$V(D) = \frac{D^2}{r}, \quad W(D) = \frac{2D}{r} \tag{21}$$

The energy balance of Equation (8) gives the depth, which grows with the square root of gradient times national flexibility:

$$D = \sqrt{rE} \tag{22}$$

The local charging peak is the share times the depth:

$$\Lambda_v = s\sqrt{rE} \tag{23}$$

This is the central result of the simplified model: the local charging peak is the share of the grid times the square root of gradient times national flexibility.

The new coincident peak follows from Equation (11):

$$C_{endo}^* = B_v + s\sqrt{rE} = B_v + \sqrt{r \cdot s \cdot E_L} \tag{24}$$

Equation (24) is the result of this paper in its simplest form. Four parameters determine the new local peak: the background load in the hour of the lowest national residual load, $B_v$; the gradient of the residual load trough, $r$; the share of the local fleet in national flexibility, $s$; and the national flexibility, $E$, or, equivalently, the local battery energy $E_L = sE$. The background load enters additively; the other three enter as the square root of their product. Nothing else enters. The level and the shape of the price function cancel out. The rated power of the batteries, their number by itself, the round-trip efficiency, the network charge and the arbitrage wedge do not appear: they decide whether the energy-constrained regime applies, through Equation (26), not how high the peak is. Whether the rest of the national flexibility is located in other low-voltage grids or at the transmission level does not matter either, as long as it follows the same arbitrage logic. Under exogenous prices the counterpart is Equation (5): the new peak is the background load in the cheapest hour plus the rated power of the local fleet, two parameters, and the battery energy sets only the duration of the peak.

The overshoot condition of Equation (12) becomes a threshold on local battery energy:

$$C^*_{endo} > B^{max} \quad \Leftrightarrow \quad E_L > \frac{(B^{max} - B_v)^2}{r \cdot s} \tag{25}$$

The threshold in Equation (25) holds the share constant, so that the national fleet grows with the local one. If instead the national flexibility is fixed and only the local fleet grows, as in a frontrunner grid, the threshold is headroom times effective charging duration, $E_L > (B^{max} - B_v) \cdot W_{eff}$, where the effective charging duration of Equation (10) is $W_{eff} = V(D)/D = \sqrt{E/r}$: it depends on national flexibility and gradient alone. The regime boundary of Equation (15) becomes a condition on national flexibility alone:

$$W_{eff} = \sqrt{\frac{E}{r}} \geq \tau \quad \Leftrightarrow \quad E \geq r\,\tau^2 \tag{26}$$

## 5.2. Scaling

Equations (23) and (24) state the scaling laws. The local charging peak grows with the square root of local battery energy, with the square root of the share, and with the square root of the gradient. Doubling the local fleet at constant share, which is a uniform national rollout, raises the peak by a factor of 1.41; doubling the share at constant local fleet, which is a frontrunner grid, by the same factor; doubling both, which is large batteries in one grid only, by a factor of two. The share alone is not sufficient, because the depth depends on the absolute national flexibility; the V-shaped form is what collapses both into a single product. The gradient is the sum of the ramps of load and wind: constant wind lowers the whole trough without changing its slope, a short wind peak at the national load minimum makes the trough pointed. The most dangerous night combines a pointed trough with a bottom early in the evening, when the background load is high.

For troughs of other shapes the exponent of the law changes. If the residual load rises from its minimum as a power $\alpha$ of time, $R^0_t = R_{min} + \kappa|t - t_0|^\alpha$, the volume is proportional to $D^{(\alpha+1)/\alpha}$ and the depth to $E^{\alpha/(\alpha+1)}$. The elasticity of the local peak with respect to national flexibility is one half for the V-shaped trough, two thirds for a parabolic trough and one for a flat-bottomed trough of fixed width, in which charging is simply the local energy divided by the width. The critical nights of 2025 are close to the V-shaped case, which is what Table 3 shows.

## 5.3. Comparison with simulations

Table 2 collects the parameters of the base case of Hirth (2026b) and the quantities the model derives from them. Under exogenous prices the model reproduces the simulated peak exactly. Under endogenous prices, with all flexibility in the denominator, the model gives 136 kW versus 163 kW in the simulation; with the home fleet alone, which the cost ordering of the simulation justifies, 159 kW. The threshold at constant share is 890 kWh of local energy, or 111 batteries; at fixed national flexibility it is 596 kWh, or 75 batteries; with the home fleet alone in the denominator, 455 kWh, or 57 batteries, which matches the simulation that stays just below the old peak at 50 batteries.

Table 2. Parameters and results of the V-shaped special case, base case of Hirth (2026b), 50 % of households with a battery

| Quantity | Value | Source |
|---|---|---|
| Old import peak $B^{max}$ | 169 kW | simulation |
| Background load in the cheapest hour $B_c$ | 78 kW (10 January, 4 a.m.) | simulation |
| Background load at the lowest residual load $B_v$ | 68 kW (10 December, 5 a.m.) | simulation |
| Batteries $X$, rated power $p$, energy $e$ | 50, 4 kW, 8 kWh | assumption |
| Rated power in the grid $Xp$, local energy $E_L$ | 200 kW, 400 kWh | derived |
| National flexibility $E$ (home fleet / all) | 76 / 136 GWh | assumption |
| Share $s$ (home fleet / all) | $5.3 \cdot 10^{-6}$ / $2.9 \cdot 10^{-6}$ | derived |
| Flank gradient $r$, critical night (median of winter nights) | 3.9 (1.9) GW/h | residual load 2025 |
| Depth $D$ (home fleet / all) | 17 / 23 GW | Eq. (22) |
| Effective charging duration $W_{eff}$ (home fleet / all) | 4.5 / 5.9 h | Eq. (26) |
| Full-load duration $\tau$ | 2.2 h | derived |
| Exogenous peak $C^*_{exo}$ | 278 kW (simulation 278) | Eq. (5) |
| Exogenous threshold | 23 batteries | Eq. (6) |
| Local charging peak $\Lambda_v$ (home fleet / all) | 91 / 68 kW (simulation 93) | Eq. (23) |
| Endogenous peak $C^*_{endo}$ (home fleet / all) | 159 / 136 kW (simulation 163) | Eq. (24) |
| Endogenous threshold, constant share (home fleet / all) | 57 / 111 batteries | Eq. (25) |
| Endogenous threshold, fixed national flexibility (home fleet / all) | 57 / 75 batteries | Section 5.1 |
| Regime boundary $r\tau^2$ | 19 GWh | Eq. (26) |

The flank gradient of the critical night, measured as depth squared over volume at 17 GW depth, is 3.9 GW/h, against a median of 1.9 GW/h over all winter nights; the fleet picks the steepest deep troughs, because the spread is largest there. Table 3 reports the ratio of the winter-night charging peak in each variant to the base case, in the simulation and under the square-root law of Equation (23). Across battery penetration the law holds to within two percentage points. Halving the number of grids, which doubles the share, raises the simulated peak by 42 % versus 41 % in the model; three quarters of the grids give 14 versus 15 %. Halving battery energy and power gives 0.69 versus 0.71. Doubling or halving the rated power at constant energy leaves the simulated peak unchanged to the first decimal, in the same hour of the same night, which is the clearest confirmation that rated power does not bind. Doubling energy and power gives 1.57 versus 1.41: the larger fleet moves the critical night and the gradient with it.

Table 3. Winter-night charging peak of the local fleet relative to the base case: simulation of Hirth (2026b) versus the square-root law

| Variant | Simulation | Square-root law |
|---|---|---|
| 10 batteries instead of 50 | 0.43 | 0.45 |
| 25 batteries | 0.71 | 0.71 |
| 40 batteries | 0.89 | 0.89 |
| Half of the grids (share doubled) | 1.42 | 1.41 |
| Three quarters of the grids | 1.14 | 1.15 |
| 4 kWh / 2 kW | 0.69 | 0.71 |
| 16 kWh / 8 kW | 1.57 | 1.41 |
| 8 kWh / 8 kW | 1.00 | 1.00 |
| 8 kWh / 2 kW | 1.00 | 1.00 |
| No utility-scale flexibility | 1.19 | 1.00 |
| Utility-scale flexibility 20 GW | 1.04 | 1.00 |
| Non-responsive price component removed | 0.81 | 1.00 |
| Price function flatter / steeper | 1.13 / 1.10 | 1.00 |
| 50 TWh more wind | 1.45 | not computed |
| 50 TWh more solar | 1.01 | 1.00 |

The deviations are informative. Without utility-scale flexibility the simulated peak is 19 % higher: the utility-scale storage does not charge in the trough, but its evening discharge lowers the evening price and narrows the spread on which the home fleet bases its charging decision, as Section 4.3 anticipates. Without the non-responsive price component the peak is 19 % lower: that component synchronizes charging in single hours beyond what the trough implies. Halving or doubling the slope of the price function changes the peak by 10 to 13 % in both directions; the model predicts no change, and the residual effect runs through the share of the non-responsive component in the price. More wind steepens the troughs of the critical nights. More solar does not affect the winter night.

# 6. Conclusions

Whether price-responsive batteries create a new coincident peak in a low-voltage grid reduces to a comparison of two quantities. The local peak is the background load in the critical hour plus the charging of the local fleet in that hour. A rebound peak, that is a new peak above the no-battery level, occurs as soon as the fleet charging exceeds the headroom, the distance between the old peak and the background load in the critical hour.

Under exogenous prices the critical hour is the cheapest hour of the night and the fleet charges at rated power. The peak is $C^*_{exo} = B_c + Xp$: the background load $B_c$ in the cheapest hour plus the rated power $Xp$ of the $X$ batteries installed in the grid. Two parameters decide, and nothing else matters: battery energy, the price level, the size of the national fleet and the network charge, as long as it is refunded or saved, do not enter. The same holds for time-of-use tariffs, time-varying network charges and a single grid in a large market, because each of these faces a price that does not respond.

Under endogenous prices the national fleet fills the trough of residual load up to the price plateau. The peak is $C^*_{endo} = B_v + s \cdot V^{-1}(E)$: the background load $B_v$ in the hour of the lowest residual load plus the share $s$ of the grid in the national flexibility $E$ times the depth of the fill $V^{-1}(E)$, where $V$ is the volume function of the trough. For a trough with linear flanks of gradient $r$ the depth is $\sqrt{rE}$, so that the local charging peak is $\Lambda_v = s\sqrt{rE}$, Equation (23), and the peak is $C^*_{endo} = B_v + s\sqrt{rE}$. Four parameters decide: the background load at the lowest residual load, the share of the grid in national flexibility, the national flexibility itself and the gradient of the trough, which measures how pointed it is.

The shape of the price function cancels out. Any price function that increases strictly in residual load leads to this result, because a flat price means a flat residual load, and the plateau is then fixed by an energy balance in which no price appears. The result therefore does not depend on the merit order of a particular power system, only on its residual load.

Energy replaces rated power. Under endogenous prices the battery energy determines the peak; rated power is a constraint that practically never binds, because the plateau spreads charging over more hours than a battery needs to fill at rated power.

The results agree closely with the simulations of Hirth (2026b). The square-root law reproduces the simulated charging peaks within a few percent across battery penetration, local concentration and battery size, and the simulated peak does not change when rated power is doubled or halved at constant energy.

The parameters that do not affect the peak can be identified as well, and the list is long. Under endogenous prices the level, slope and shape of the price function, the rated power, the number of batteries by itself, the round-trip efficiency, the network charge and the arbitrage wedge do not enter; they decide whether the energy-constrained regime applies, not how high the peak is. Market size is irrelevant as well. Splitting a market into price zones or moving to locational marginal pricing changes neither the share nor the depth that matter, as long as the split is proportional. What matters is the share of a grid in the flexibility of its own market and the shape of that market's residual load.

The gradient of the trough is the sum of the load ramp and the wind ramp. Constant wind lowers the trough without sharpening it. The worst case is a short wind peak at night that falls on the national load minimum, or one early in the evening, when the background load is still high.

New export peaks are discharge events on price peaks, not charging events in the solar trough. Under endogenous prices they arise only with high rated power on isolated price spikes that the price feedback does not remove. They are dampened by the low PV generation in hours of high residual load, which follows from the correlation of weather across the market, and by the energy left in the battery when the price peak arrives.

Charging at the solar peak is harmless only because local PV generation runs counter to it. Where local PV is low while the national solar peak sets the price, the midday is a power-constrained event in the local grid, and its size follows from the exogenous case.

# Statements

## Acknowledgements

I thank Johanna Bronisch, Anselm Eicke and Clemens Lohr for helpful discussions.

## Competing interest

The author is managing director of Neon Neue Energieökonomik, a consultancy advising clients in the electricity sector. This paper did not receive any funding and was not commissioned by any client; no client has seen its results prior to publication.

## Funding

This research did not receive any specific grant from funding agencies in the public, commercial, or not-for-profit sectors.

## Data availability

The simulation results used for validation are those of Hirth (2026b, 2026c); the scripts that measure trough gradient and charging peaks from them are available from the author on request.

## CRediT contribution statement

Lion Hirth: Conceptualization, Methodology, Formal analysis, Writing – original draft, Writing – review & editing.